%% file: Monopole2011.tex
\documentclass[aps,prl,twocolumn,showpacs,superscriptaddress,groupedaddress]{revtex4-1}  
\pdfoutput=1           
\usepackage{graphicx}  
\usepackage{dcolumn}   
\usepackage{bm}        
\usepackage{amssymb}   

\usepackage{amsmath}
\usepackage{hyperref}

\usepackage{preprintcover}  

\PreprintCoverPaperTitle{Search for magnetic monopoles in $\sqrt{s}=7$~TeV $pp$ collisions with the ATLAS detector}  

\PreprintIdNumber{CERN-PH-EP-2012-177}  

\PreprintCoverAbstract{
This Letter presents a search for magnetic monopoles with the ATLAS
detector at the CERN Large Hadron Collider using an integrated
luminosity of 2.0~fb$^{-1}$ of $pp$ collisions recorded at a
center-of-mass energy of $\sqrt{s}=7$~TeV. No event is found in the
signal region, leading to an upper limit on the production cross
section at 95\% confidence level of $1.6/\epsilon$~fb for Dirac
magnetic monopoles with the minimum unit magnetic charge and with mass
between 200~GeV and 1500~GeV, where $\epsilon$ is the monopole
reconstruction efficiency.  The efficiency $\epsilon$ is high and
uniform in the fiducial region given by pseudorapidity $|\eta|<1.37$
and transverse kinetic energy $600$--$700<E^{\rm
  kin}\sin\theta<1400$~GeV. The minimum value of $700$~GeV is for
monopoles of mass 200~GeV, whereas the minimum value of $600$~GeV is
applicable for higher mass monopoles. Therefore, the upper limit on
the production cross section at 95\% confidence level is 2~fb in this
fiducial region.  Assuming the kinematic distributions from Drell-Yan
pair production of spin-1/2 Dirac magnetic monopoles, the efficiency
is in the range 1\%--10\%, leading to an upper limit on the cross
section at 95\% confidence level that varies from 145~fb to 16~fb for
monopoles with mass between 200~GeV and 1200~GeV. This limit is weaker
than the fiducial limit because most of these monopoles lie outside
the fiducial region.
}  

\PreprintJournalName{Physical Review Letters}  

\begin{document}

\title{Search for magnetic monopoles in $\sqrt{s}=7$~TeV $pp$ collisions with the ATLAS detector}

\author{The ATLAS Collaboration}

\begin{abstract}
This Letter presents a search for magnetic monopoles with the ATLAS
detector at the CERN Large Hadron Collider using an integrated
luminosity of 2.0~fb$^{-1}$ of $pp$ collisions recorded at a
center-of-mass energy of $\sqrt{s}=7$~TeV. No event is found in the
signal region, leading to an upper limit on the production cross
section at 95\% confidence level of $1.6/\epsilon$~fb for Dirac
magnetic monopoles with the minimum unit magnetic charge and with mass
between 200~GeV and 1500~GeV, where $\epsilon$ is the monopole
reconstruction efficiency.  The efficiency $\epsilon$ is high and
uniform in the fiducial region given by pseudorapidity $|\eta|<1.37$
and transverse kinetic energy $600$--$700<E^{\rm
  kin}\sin\theta<1400$~GeV. The minimum value of $700$~GeV is for
monopoles of mass 200~GeV, whereas the minimum value of $600$~GeV is
applicable for higher mass monopoles. Therefore, the upper limit on
the production cross section at 95\% confidence level is 2~fb in this
fiducial region.  Assuming the kinematic distributions from Drell-Yan
pair production of spin-1/2 Dirac magnetic monopoles, the efficiency
is in the range 1\%--10\%, leading to an upper limit on the cross
section at 95\% confidence level that varies from 145~fb to 16~fb for
monopoles with mass between 200~GeV and 1200~GeV. This limit is weaker
than the fiducial limit because most of these monopoles lie outside
the fiducial region.
\end{abstract}

\maketitle

Magnetic monopoles have long been the subject of dedicated search
efforts for three main reasons: their introduction into the theory of
electromagnetism would restore the symmetry between electricity and
magnetism in Maxwell's equations; their existence would explain the
quantization of electric charge~\cite{Dirac}; and they appear in many
grand unified theories~\cite{Hooft}.  However, to date no experimental
evidence of a magnetically charged object exists.

Recent searches for magnetic monopoles from astrophysical
sources~\cite{MACRO,SLIM,Baikal,RICE,IceCube,ANITA,ANTARES} are
complemented by searches at colliders~\cite{d0,h1,cdf,opal,moedal}.
This Letter describes a search for magnetic monopoles in
proton--proton collisions recorded at a center-of-mass energy of
$\sqrt{s}=7$~TeV using the ATLAS detector at the CERN Large Hadron
Collider (LHC).

The Dirac quantization condition~\cite{Dirac}, given in Gaussian
units, leads to a prediction for the minimum unit magnetic charge $g$:
\begin{equation}
\frac{ge}{\hbar c}=\frac{1}{2}\Rightarrow \frac{g}{e}=\frac{1}{2\alpha_e}\approx 68.5,
\label{eqn:gcharge}
\end{equation}
where $e$ is the unit electric charge and $\alpha_e$ is the fine
structure constant. With the introduction of a magnetic monopole, the
duality of Maxwell's equations implies a magnetic
coupling~\cite{MonoNIM}
\begin{equation}
\alpha_m = \frac{(g\beta)^2}{\hbar c} = \frac{1}{4\alpha_e}\beta^2, \\
\label{eqn:magcoupl}
\end{equation}
where $\beta=v/c$ is the monopole velocity. For relativistic
monopoles, $\alpha_m$ is very large, precluding any perturbative
calculation of monopole production processes. Therefore, the main
result of this analysis is a fiducial cross-section limit for Dirac
monopoles of magnetic charge $g$ derived without assuming a particular
production mechanism.  A cross-section limit assuming the kinematic
distributions from Drell-Yan monopole pair production is also
provided.

Monopoles are highly ionizing particles, interacting with matter like
an ion of electric charge $68.5e$, according to Eq.~\ref{eqn:gcharge}.
The high stopping power of the monopole ionization~\cite{ahlen}
results in the production of a large number of $\delta$-rays. These
energetic ``knock-on'' electrons emitted from the material carry away
energy from the monopole trajectory and further ionize the medium. In
the mass and energy regime of this study, the $\delta$-rays have
kinetic energies ranging from 1~MeV to a maximum of $\sim \!100$~MeV.
The secondary ionization by these $\delta$-rays represents a
significant fraction of the ionization energy loss of the magnetic
monopole~\cite{ahlen}.  The dominant energy loss mechanism for
magnetic monopoles in the mass and energy range considered herein is
ionization~\cite{craigie,giacomelli,ahlen}. Furthermore, the monopole
ionization is independent of the monopole speed $\beta$ to first
order, in contrast to the ionization of electrically charged
particles.

In the ATLAS detector~\cite{ATLAS,coord}, the monopole signature can
be easily distinguished using the transition radiation tracker (TRT)
in the inner detector and the liquid argon (LAr) sampling
electromagnetic (EM) calorimeter. The TRT is a straw-tube tracker that
comprises a barrel ($|\eta|<1.0$) with 4~mm diameter straws oriented
parallel to the beam-line, and two endcaps ($0.8<|\eta|<2.0$) with
straws orientated radially. A minimum ionizing particle deposits $\sim
\!\!2$~keV of energy in a TRT straw. Energy deposits in a TRT straw
greater than 200~eV (called ``low-threshold hits'') are used for
tracking, while those that exceed 6~keV (called ``high-threshold
hits'') typically occur due to the transition radiation emitted by
highly relativistic electrons when they penetrate the radiator layers
between the straws.  As a result, an electron of energy 5~GeV or above
has a 20\% probability of producing a high-threshold hit in any straw
it traverses. The high-threshold hits can also indicate the presence
of a highly ionizing particle. A 2~T superconducting solenoid magnet
surrounds the inner detector. The LAr barrel EM calorimeter lies
outside the solenoid in the $|\eta|<1.5$ region. It is divided into
three shower-depth layers and comprises accordion-shaped electrodes
and lead absorbers.  The cell granularity in the second layer is
$\Delta \eta \times \Delta \phi = 0.025\times0.025$. The
characteristic signature of magnetic monopoles in ATLAS is a large
localized energy deposit in the LAr EM calorimeter (EM cluster) in
conjunction with a region of high ionization density in the TRT.  A
search for particles with large electric charge, which yield a similar
signature, was performed previously~\cite{Qball} and production
cross-section limits for such particles were set~\cite{Mermod}.

The trajectory of an electrically neutral magnetic monopole in the
inner detector is straight in the $r$--$\phi$ plane and curved in
$r$--$z$.  The behavior of magnetic monopoles in the ATLAS detector is
described by a {\sc Geant4}~\cite{geant4} simulation~\cite{ATLSIM},
which includes the equations of motion, the ionization, the
$\delta$-ray production and a modified Birks' Law~\cite{Birks} to
model recombination effects in LAr due to highly ionizing
particles~\cite{HIPBirksNIM}. Equation~5.5 in Ref.~\cite{ahlen} gives
the $\delta$-ray production cross section and Eq.~5.7 describes the
derivation of the magnetic monopole ionization; both equations are
implemented in {\sc Geant4}.

Simulated Monte Carlo (MC) single-monopole samples are used to
determine the efficiency as a function of the transverse kinetic
energy $E_{\rm T}^{\rm kin}=E^{\rm kin}\sin\theta$ and pseudorapidity
$\eta$ for various monopole masses.  For the Drell-Yan process, it is
assumed that spin-1/2 magnetic monopoles are produced in pairs from
the initial $pp$ state via quark-antiquark annihilation into a virtual
photon. MadGraph~\cite{MadGraph} is used to model this process by
assuming leading-order Drell-Yan heavy lepton pair production but
making the replacement $e\rightarrow g \beta$ to reflect the magnetic
coupling in Eq.~\ref{eqn:magcoupl}.  In the absence of a consistent
theory describing the coupling of the monopole to the $Z$ boson, such
a coupling is set to zero in the MadGraph model.  In the Drell-Yan
samples, the CTEQ6L1~\cite{cteq} parton distribution functions are
used and {\sc Pythia} version 6.425~\cite{pythia} is used for the
hadronization and the underlying event. Only Drell-Yan monopoles with
transverse momentum $p_{\rm T}>200$~GeV are processed by the
simulation since lower $p_{\rm T}$ monopoles fail to reach the
calorimeter.  For all the simulated samples, both the monopoles and
the antimonopoles are assumed to be stable and all final-state
particles are processed by the simulation of the ATLAS detector.
Additional $pp$ collisions in each event are simulated according to
the distribution of $pp$ interactions per bunch crossing in the
selected data period.

A simple algorithm is used to preselect events with monopole
candidates for further study. Monopoles with $E_{\rm T}^{\rm kin}$
above approximately 500~GeV traverse the inner detector and penetrate
to the LAr calorimeter, depositing most of their energy there. Only
one third of the deposited energy is recorded due to the recombination
effects in LAr~\cite{HIPBirksNIM}.  Lacking a dedicated monopole
trigger, only events collected with a single-electron trigger with
transverse energy threshold $E_{\rm T}>60$~GeV are considered. This
trigger requires a track in the inner detector within $|\Delta
\eta|<0.01$ and $|\Delta \phi|<0.02$ of the LAr energy deposit.
Monopoles that fulfill the 60~GeV energy requirement travel fast
enough to satisfy the tracking and timing requirements of the
trigger. Very high energy monopoles (i.e., those with $E_{\rm T}^{\rm
  kin}\gtrsim 1400$--$1900$~GeV, where the value of 1400~GeV is for
monopoles of mass~1500 GeV and the value of $1900$~GeV is for
monopoles of mass 200~GeV) exit the EM calorimeter and are rejected by
a veto on hadronic energy that is intrinsic to the single-electron
trigger.  This trigger was operational during the first six months of
2011 data-taking and recorded an integrated luminosity of
2.0~fb$^{-1}$, defining the dataset used for this search.

The reconstructed EM cluster is then required to have $E_{\rm
  T}>65$~GeV and $|\eta|<1.37$. The trigger efficiency is independent
of $E_T$ for $E_{\rm T}>65$~GeV, motivating the former requirement.
The $\eta$ requirement ensures that the EM cluster is in the barrel
region of the LAr calorimeter, where the two-dimensional spatial
resolution is uniform.  If two or more EM clusters in an event satisfy
these criteria, only the cluster with the highest energy is considered
as a monopole candidate.

In the barrel region, the monopole typically traverses 35 TRT straws
and its high ionization ensures that most of these register
high-threshold hits. Furthermore, as each $\delta$-ray produced by the
monopole ionization deposits $\sim \!2$~keV in a straw, the combined
energy deposited by multiple $\delta$-rays crossing a single TRT straw
gives rise to additional high-threshold hits.  The large number of
$\delta$-rays bend in the 2~T magnetic field in the $r$--$\phi$ plane;
therefore, the monopole trajectory appears as a $\sim \!1$-cm-wide
swath of high-threshold TRT hits.  The fraction of TRT hits that
exceed the high threshold in the vicinity of the path of an ionizing
particle is therefore a powerful discriminator between the monopole
signal and the background. The $\phi$ position of the EM cluster is
used to define a road of width $\Delta \phi=\pm0.05$~rad from the
beam-line to the cluster. At least twenty high-threshold TRT hits must
be present in the road.  In addition, at least 20\% of the TRT hits in
the road must be high-threshold hits.

After the preselection, a more refined TRT hit counting algorithm is
used to distinguish the signal from the backgrounds. A histogram with
a bin width of 0.8~mrad is filled with the $\phi$ distribution of the
high-threshold hits in the previously defined road. The location of
the highest bin is used to calculate the center of a new road.  In the
TRT barrel, a rectangular road of $\pm4$~mm in the $r$--$\phi$ plane
is used and the hits are counted.  In the TRT endcap, a wedge-shaped
road of width $\Delta \phi=\pm0.006$~rad is used.  These roads are
wide enough to encompass two neighboring straws, taking into account
the monopole trajectory and the associated $\delta$-rays, but
sufficiently narrow to ensure that the fraction of hits that exceed
the high threshold, $f_{\rm HT}$, is insensitive to the presence of
neighboring tracks.  In the barrel region, the number of hits in the
road is required to be greater than 54.  An $\eta$-dependent
requirement on the number of hits in the road is applied in the endcap
and barrel--endcap transition region to account for the TRT geometry.

Energy loss by bremsstrahlung and $e^+e^-$ pair production is
negligible for magnetic monopoles in the mass and energy range
considered herein. Therefore, magnetic monopoles give rise to a narrow
ionization energy deposit in the LAr calorimeter, the size of which
provides another powerful discriminator of the monopole signal from
backgrounds such as electrons and photons, which induce an EM shower
via bremsstrahlung and $e^+e^-$ pair production.  The variable used is
$\sigma_{\rm R}$, the energy-weighted two-dimensional $\eta$--$\phi$
cluster dispersion in the second layer of the EM calorimeter, which
has the highest two-dimensional spatial resolution. The dispersion
$\sigma_{\rm R}$ is calculated from the energies deposited in a
$3\times7$ array of cells centered around the most energetic cell of
the EM cluster: $\sigma_{\rm R} = \sqrt{\sigma_\phi^2+\sigma_\eta^2}$,
where $\sigma_\phi^2 = \Sigma\left(E_i \delta \phi_i^2\right)/\Sigma
E_i -\left[\Sigma \left(E_i\delta\phi_i\right)/\Sigma E_i\right]^2$,
$\delta \phi_i$ is the deviation in $\phi$ between cell~$i$ and the
most energetic cell and $E_i$ is the energy of cell~$i$;
$\sigma_\eta^2$ is defined similarly.

\begin{figure}
  \centering
  \includegraphics[width=0.45\textwidth]{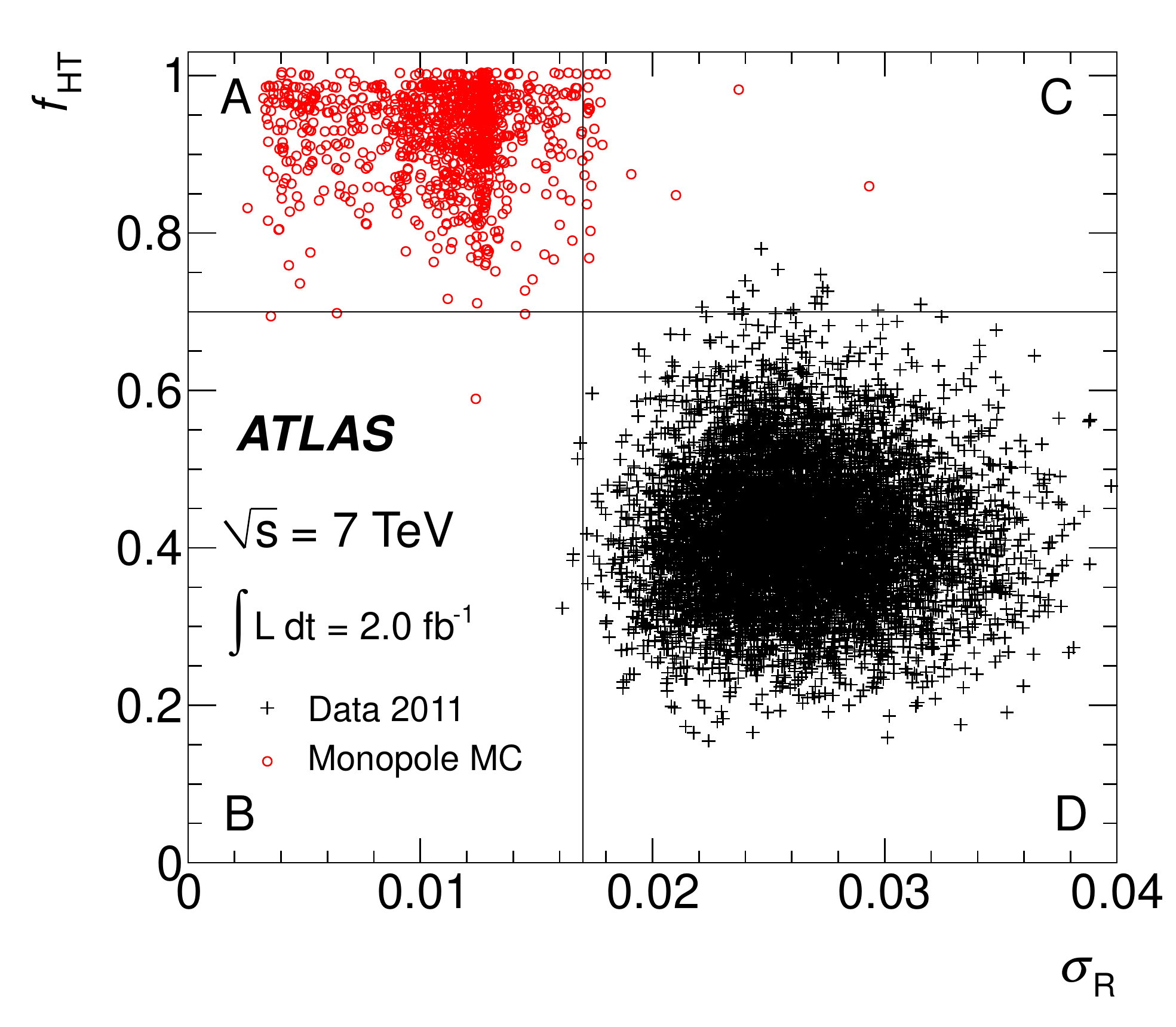}
  \caption{High-threshold TRT hit fraction, $f_{\rm HT}$, versus EM
    cluster dispersion, $\sigma_{\rm R}$. The circles represent 1000
    simulated single monopoles with mass 800~GeV.  The crosses
    represent ATLAS data. The regions marked A-D are discussed in the
    text.}
  \label{fig:htfrac_deltaR}
\end{figure}

The high-threshold TRT hit fraction, $f_{\rm HT}$, and the cluster
dispersion, $\sigma_{\rm R}$, are thus chosen as the distinguishing
variables between the signal and background, and are shown in
Fig.~\ref{fig:htfrac_deltaR}.  The main physics background sources are
high-energy electrons, photons and jets, which exhibit no correlation
between these variables in simulated processes. The background and
monopole MC samples are used to define an approximate signal region.
Then ($\sigma_{\rm R}$, $f_{\rm HT}$) parameter pairs are generated by
randomly sampling the one-dimensional $\sigma_{\rm R}$ and $f_{\rm
  HT}$ distributions for data outside this approximate signal
region. The borders of the signal region are tuned for maximal
significance of observation of three signal events by replacing the
background MC events with these parameter pairs.  The final signal
region A is defined by $\sigma_{\rm R} \leq 0.017$ and $f_{\rm
  HT}>0.7$.

\begin{figure}
  \centering
  \includegraphics[width=0.45\textwidth]{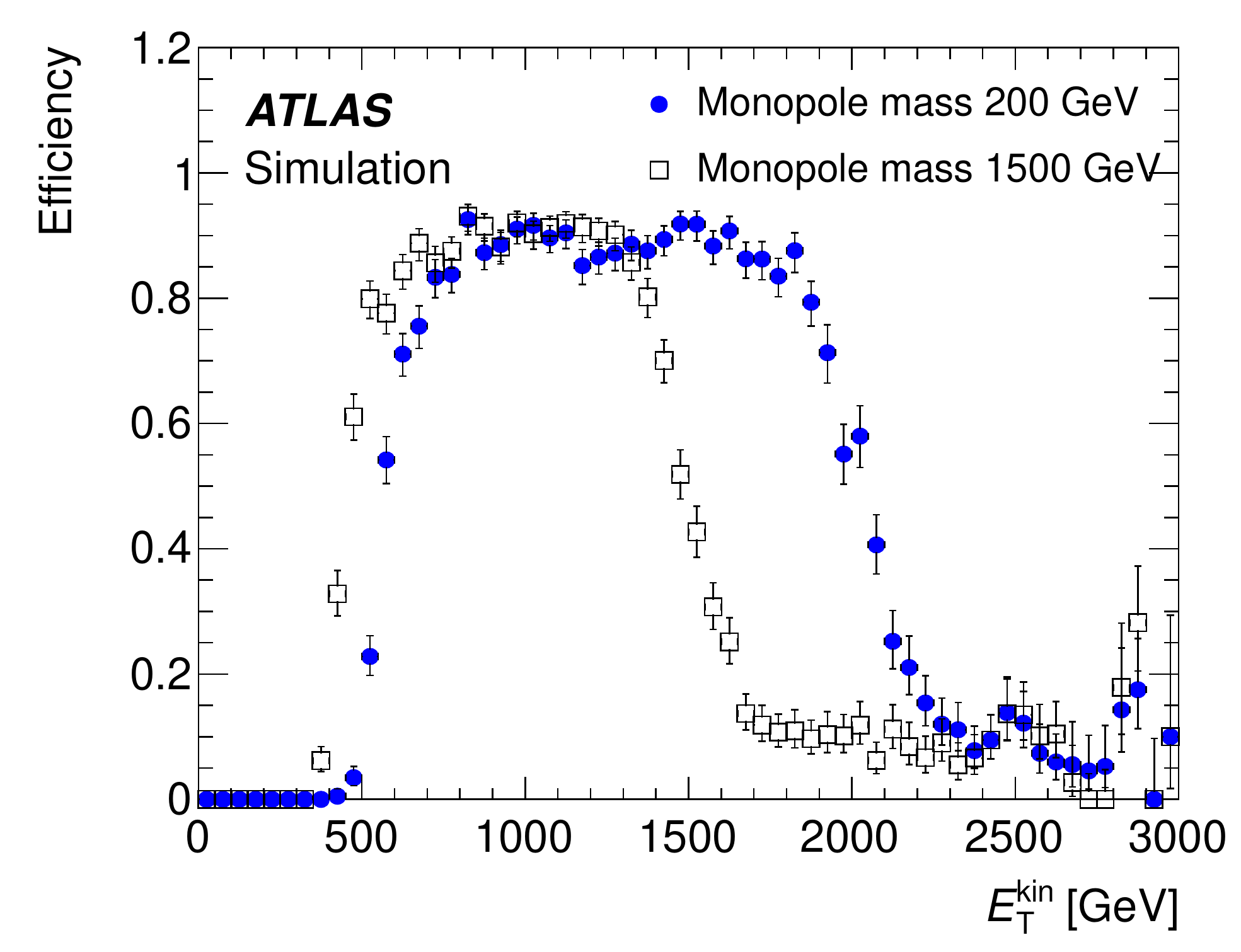}
  \caption{Efficiency versus $E_{\rm T}^{\rm kin}$, averaged over
    $|\eta|<1.37$, for single monopoles of mass 200~GeV and mass
    1500~GeV.}
  \label{fig:eksinthetaeff}
\end{figure}

The efficiencies, which include trigger, reconstruction and selection
effects, in the two-dimensional $E_{\rm T}^{\rm kin}$ versus $\eta$
plane are obtained from the simulated single-monopole samples.  A
fiducial region for each monopole mass is defined by the $E_{\rm
  T}^{\rm kin}$ range in which the efficiency is 0.80 or higher in the
$|\eta|<1.37$ region.  Figure~\ref{fig:eksinthetaeff} shows the
efficiency versus $E_{\rm T}^{\rm kin}$, averaged over $|\eta|<1.37$.
For monopoles with a mass of 200~GeV, the minimum transverse kinetic
energy $(E_{\rm T}^{\rm kin})_{\rm min}$ where the efficiency rises
above 0.80 is 700~GeV. For monopoles with a mass between 500~GeV and
1500~GeV, $(E_{\rm T}^{\rm kin})_{\rm min}$ is 600~GeV. Monopoles with
lower $E_{\rm T}^{\rm kin}$ fail to penetrate to the EM calorimeter
and therefore do not satisfy the trigger requirements. Monopoles with
very high $E_{\rm T}^{\rm kin}$ exit the EM calorimeter and are
rejected by the hadronic veto of the electron trigger. A common upper
value of $E_{\rm T}^{\rm kin}=1400$~GeV is used for the fiducial
region of all monopole masses.  As the minimum efficiency is 0.80 in
the fiducial region, a common value of 0.80 is used in the
determination of the upper cross-section limit.

The efficiencies can be under- or over-estimated for several reasons.
These effects are described below and the relative systematic
uncertainties for each effect are given.  1) Cross-talk in the second
EM layer in the $\phi$ direction is not modeled in the simulation.
The energy is reweighted assuming 1.8\% cross-talk~\cite{crosstalk}
and the cluster dispersion, $\sigma_{\rm R}$, recomputed. The
efficiency is reduced and the resulting relative shift of $-1.7$\% for
single monopoles is taken as a one-sided uncertainty.  2) The
simulation underestimates the TRT occupancy in the data by up to 20\%;
therefore, the number of low-threshold hits (those unlikely to come
from the monopole or related $\delta$-rays) is increased by 20\%.  The
resulting relative uncertainty is $-1.3$\%.  3) The modification to Birks' Law
is varied between its upper and lower systematic
uncertainties~\cite{HIPBirksNIM}, yielding a relative uncertainty of $+1.8$\%
and $+1.5$\%, respectively.  4) The production of $\delta$-rays is
varied by 3\%~\cite{ahlen} and the resulting uncertainty is
negligible.  5) The {\sc Geant4} ``range cut''~\cite{geant4} controls
the minimum kinetic energy threshold below which $\delta$-rays are not
propagated explicitly. This parameter is reduced from 50~$\mu$m to
25~$\mu$m in the TRT simulation.  The resulting relative uncertainty is
$+0.14$\%.  6) The material in the inner detector, in the barrel
cryostat and in between the cryostat and the first layer of the EM
calorimeter is increased by 5\%, 10\% and 5\%, respectively, in the
simulation~\cite{material}.  The resulting $-0.74$\% relative uncertainty is
taken as symmetric.  Including an uncertainty of $\sim \!\!1.7$\% to
account for the limited number of MC events, the total upper and lower
relative uncertainties on the efficiency for single monopoles are
$+2.6$\% and $-2.8$\%, respectively.

The efficiencies to reconstruct at least one of the monopoles in the
pairs produced with Drell-Yan kinematic distributions are given in
Table~\ref{tbl:DYsyst} for each mass. Only masses up to 1200~GeV are
considered, taking into account the phase space limitations for pair
production. The total relative uncertainties, which reflect the same
systematic variations described above, are also given.  The
efficiencies and their associated systematic uncertainties reflect
large losses due to acceptance, since many Drell-Yan monopoles have
insufficient energy to reach the calorimeter.

\begin{table}[htb]
\centering
\caption{Efficiencies and their relative uncertainties in percent for
  Drell-Yan pair-produced monopoles of various masses.}
\begin{tabular}{crrrrr}
\hline
\hline
Mass (GeV) & 200  & 500  & 800  & 1000  & 1200\\
\hline
Efficiency & 0.011 & 0.048  & 0.081 & 0.095 & 0.095 \\
\hline
Relative uncertainty & & & & & \\
Upper (\%) & $+32$ & $+24$ & $+22$ & $+23$ & $+20$\\
Lower (\%) & $-36$ & $-23$ & $-22$ & $-25$ & $-25$\\
\hline
\hline
\end{tabular}
\label{tbl:DYsyst}
\end{table}

The background in the signal region is predicted directly from the
data.  The two-dimensional plane in Fig.~\ref{fig:htfrac_deltaR} is
divided into quadrants, one of which is dominated by signal (region
A), and three others that are occupied mainly by background (regions
B, C and D). The ratio of background events in signal region A to
events in background region B is expected to be the same as the ratio
of background events in regions C to D.  This assumption is
incorporated into a maximum likelihood fit to determine the estimated
numbers of signal and background events in signal region A. The inputs
to the fit include the observed event yields in quadrants A through D,
which are 0, 5, 16 and 7001, respectively, the efficiencies and
associated systematic uncertainties that have already been discussed,
and the integrated luminosity and its 3.7\%
uncertainty~\cite{lumi}. For each monopole mass, the rate of
appearance of signal events in quadrants B and C, as predicted by the
simulation, is also taken into account. According to the simulation,
no signal event appears in quadrant D for any monopole mass.  The
fit predicts $0.011\pm0.007$ background events in the signal region.

\begin{figure}[!htb]
  \centering
  \includegraphics[width=0.45\textwidth]{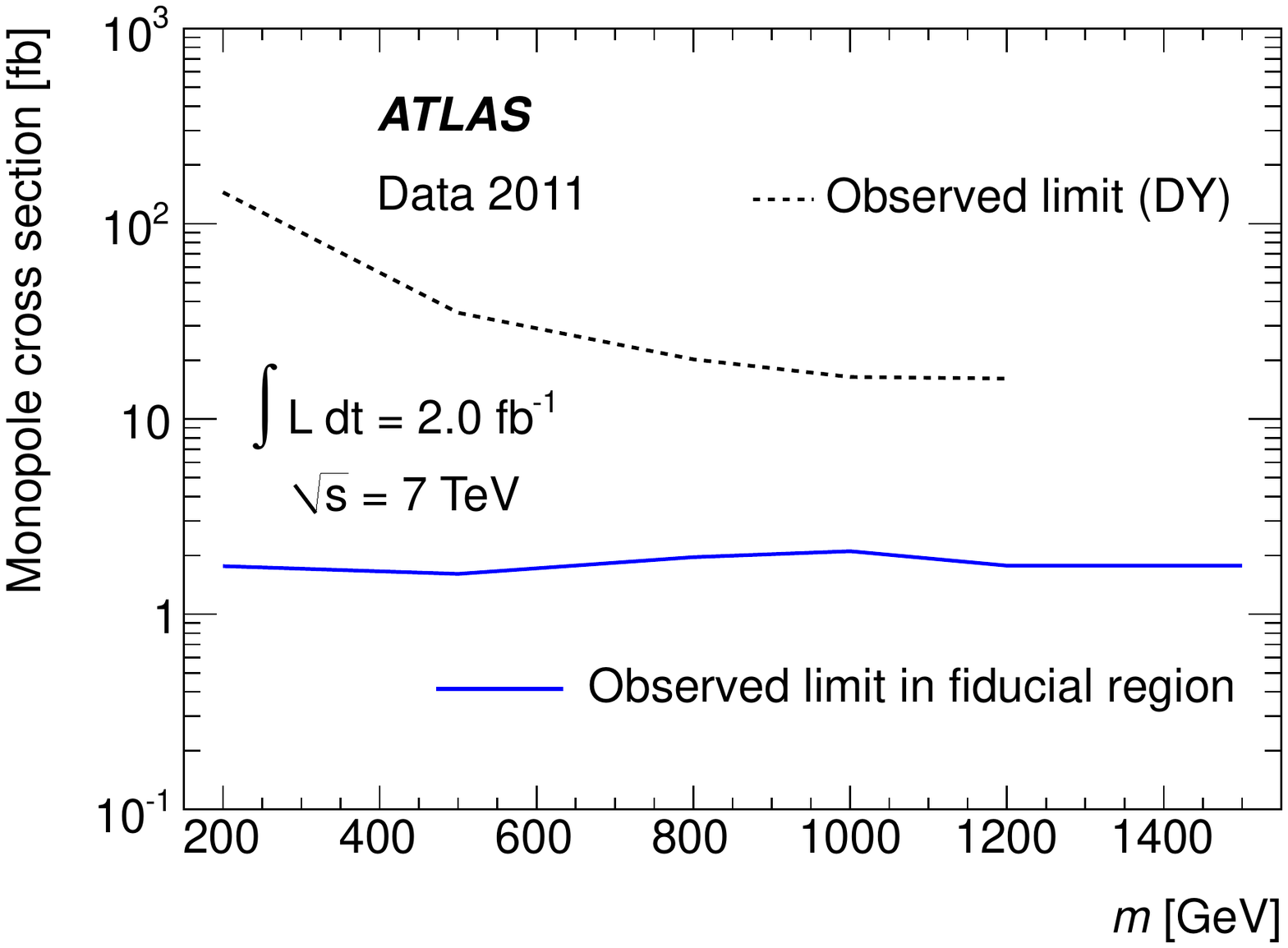}
  \caption{Upper limits on the monopole production cross sections at
    95\% confidence level.  The solid line is the limit for single
    monopoles in the fiducial region and the dashed line is the limit
    assuming the kinematic distributions from Drell-Yan (DY) monopole
    pair production.}
  \label{fig:xs}
\end{figure}

Using the results of the maximum likelihood fit, the upper limits on
the production cross sections at 95\% confidence level are calculated
using the profile likelihood ratio as a test
statistic~\cite{RooStats}. The results are extracted using the $CL_s$
method~\cite{Read}.  The cross section limits can be expressed as a
function of the efficiency, $\epsilon$, which is shown in
Fig.~\ref{fig:eksinthetaeff} for single monopoles and given in
Table~\ref{tbl:DYsyst} for Drell-Yan pair-produced monopoles.  The
upper limit on the production cross section at 95\% confidence level
is found to be $1.6/\epsilon$~fb for Dirac magnetic monopoles with the
minimum unit magnetic charge and with mass between 200~GeV and
1500~GeV.  Assuming the kinematic distributions from Drell-Yan pair
production of spin-1/2 Dirac magnetic monopoles, this translates to an
upper limit on the cross section at 95\% confidence level that varies
from 145~fb to 16~fb for monopoles with mass between 200~GeV and
1200~GeV, as shown in Fig.~\ref{fig:xs}. Since the number of expected
background events is very small and no event is observed in the signal
region, only the observed limits are shown.  To compare with previous
experiments that have provided lower mass limits on spin-1/2 Dirac
magnetic monopoles by assuming Drell-Yan pair production, such an
approach would yield a lower mass limit of 862~GeV in the present
search~\cite{DYsigma}.

The monopole reconstruction efficiency is high and uniform in the
fiducial region given by pseudorapidity $|\eta|<1.37$ and transverse
kinetic energy $(E_{\rm T}^{\rm kin})_{\rm min}<E^{\rm
  kin}\sin\theta<1400$~GeV, where $(E_{\rm T}^{\rm kin})_{\rm min}$ is
600~GeV for monopoles with a mass between 500~GeV and 1500~GeV. For
monopoles with a mass of 200~GeV, $(E_{\rm T}^{\rm kin})_{\rm
  min}=700$~GeV.  Therefore, the upper limit on the production cross
section at 95\% confidence level is 2~fb, as shown in
Fig.~\ref{fig:xs}, for Dirac magnetic monopoles with the minimum unit
magnetic charge and with mass between 200~GeV and 1500~GeV in this
fiducial region.  The fluctuations of the observed limit in the
fiducial region originate from variations of the nuisance parameters
used in the profile likelihood ratio.

These results extend the upper limits on the production cross section
for monopoles in this mass region established by preceding
experiments. This is the first direct collider search that yields
cross-section constraints on magnetic monopoles with masses greater
than 900~GeV.

\begin{acknowledgments}


We thank CERN for the very successful operation of the LHC, as well as
the support staff from our institutions without whom ATLAS could not
be operated efficiently. We thank J.~Apostolakis, V.~Ivanchenko,
M.~Horbatsch and V.~Sanz for helpful discussions.

We acknowledge the support of ANPCyT, Argentina; YerPhI, Armenia; ARC,
Australia; BMWF, Austria; ANAS, Azerbaijan; SSTC, Belarus; CNPq and FAPESP,
Brazil; NSERC, NRC and CFI, Canada; CERN; CONICYT, Chile; CAS, MOST and NSFC,
China; COLCIENCIAS, Colombia; MSMT CR, MPO CR and VSC CR, Czech Republic;
DNRF, DNSRC and Lundbeck Foundation, Denmark; EPLANET and ERC, European Union;
IN2P3-CNRS, CEA-DSM/IRFU, France; GNAS, Georgia; BMBF, DFG, HGF, MPG and AvH
Foundation, Germany; GSRT, Greece; ISF, MINERVA, GIF, DIP and Benoziyo Center,
Israel; INFN, Italy; MEXT and JSPS, Japan; CNRST, Morocco; FOM and NWO,
Netherlands; RCN, Norway; MNiSW, Poland; GRICES and FCT, Portugal; MERYS
(MECTS), Romania; MES of Russia and ROSATOM, Russian Federation; JINR; MSTD,
Serbia; MSSR, Slovakia; ARRS and MVZT, Slovenia; DST/NRF, South Africa;
MICINN, Spain; SRC and Wallenberg Foundation, Sweden; SER, SNSF and Cantons of
Bern and Geneva, Switzerland; NSC, Taiwan; TAEK, Turkey; STFC, the Royal
Society and Leverhulme Trust, United Kingdom; DOE and NSF, United States of
America.

The crucial computing support from all WLCG partners is acknowledged
gratefully, in particular from CERN and the ATLAS Tier-1 facilities at
TRIUMF (Canada), NDGF (Denmark, Norway, Sweden), CC-IN2P3 (France),
KIT/GridKA (Germany), INFN-CNAF (Italy), NL-T1 (Netherlands), PIC (Spain),
ASGC (Taiwan), RAL (UK) and BNL (USA) and in the Tier-2 facilities
worldwide.

\end{acknowledgments}

\onecolumngrid
\clearpage

\input{atlas_authlist}

\end{document}

%% file: atlas_authlist.tex
\begin{flushleft}
{\Large The ATLAS Collaboration}

\bigskip

G.~Aad$^{\rm 47}$,
T.~Abajyan$^{\rm 20}$,
B.~Abbott$^{\rm 110}$,
J.~Abdallah$^{\rm 11}$,
S.~Abdel~Khalek$^{\rm 114}$,
A.A.~Abdelalim$^{\rm 48}$,
O.~Abdinov$^{\rm 10}$,
R.~Aben$^{\rm 104}$,
B.~Abi$^{\rm 111}$,
M.~Abolins$^{\rm 87}$
% ...
\bigskip

% ...
$^{10}$ Institute of Physics, Azerbaijan Academy of Sciences, Baku, Azerbaijan\\
$^{11}$ Institut de F\'{i}sica d'Altes Energies and Departament de F\'{i}sica de la Universitat Aut\`{o}noma de Barcelona and ICREA, Barcelona, Spain\\
% ...
$^{20}$ Physikalisches Institut, University of Bonn, Bonn, Germany\\
% ...
$^{47}$ Fakult\"{a}t f\"{u}r Mathematik und Physik, Albert-Ludwigs-Universit\"{a}t, Freiburg, Germany\\
$^{48}$ Section de Physique, Universit\'{e} de Gen\`{e}ve, Geneva, Switzerland\\
% ...
$^{87}$ Department of Physics and Astronomy, Michigan State University, East Lansing MI, United States of America\\
% ...
$^{104}$ Nikhef National Institute for Subatomic Physics and University of Amsterdam, Amsterdam, Netherlands\\
% ...
$^{110}$ Homer L. Dodge Department of Physics and Astronomy, University of Oklahoma, Norman OK, United States of America\\
$^{111}$ Department of Physics, Oklahoma State University, Stillwater OK, United States of America\\
% ...
$^{114}$ LAL, Universit\'{e} Paris-Sud and CNRS/IN2P3, Orsay, France\\
% ...
\end{flushleft}


%% file: Monopole2011.bbl
\begin{thebibliography}{99}
\bibitem{coord} ATLAS uses a right-handed coordinate system with its
  origin at the nominal interaction point (IP) in the center of the
  detector and the $z$-axis coinciding with the axis of the beam
  pipe. The $x$-axis points from the IP to the center of the LHC ring,
  and the $y$-axis points upward. Cylindrical coordinates ($r,\phi$)
  are used in the transverse plane, $\phi$ being the azimuthal angle
  around the beam pipe. The pseudorapidity is defined in terms of the
  polar angle $\theta$ as $\eta=-\ln \tan(\theta/2)$.
\bibitem{DYsigma}Over the range of monopole masses from 200~GeV to 1200~GeV, the leading-order Drell-Yan production cross section drops from $10^5$~fb to 1~fb, but has large theoretical uncertainties, due to the non-perturbative nature of the magnetic coupling.
\end{thebibliography}
